\documentclass[aps,pra,reprint]{revtex4-2}
\usepackage{amsmath,bm}
\usepackage{balance} 
\usepackage{graphicx}
\usepackage[T1]{fontenc}
\usepackage[colorlinks,linkcolor=blue,anchorcolor=blue,citecolor=blue]{hyperref}

\newcommand{\nn}{\nonumber}
\newcommand{\na}{\bm\nabla}
\renewcommand{\Im}{{\rm Im\,}} 
\renewcommand{\Re}{{\rm Re\,}}

\begin{document}
\title{Transverse spin and orbital angular momenta in rotating-wave structured light}
\author{Zhen-Lai Wang}
\email{wangzhenlai@hbpu.edu.cn}
\affiliation{Center for Fundamental Physics and School of Mathematics and Physics, Hubei Polytechnic University, Huangshi 435003, China}
\date{\today}
\begin{abstract}
 Within monochromatic optical fields, we demonstrate the rotating-wave structured light with wave vortex carrying an intrinsic transverse orbital angular momentum orthogonal to the propagation direction of light. Remarkably, we find that such a rotating-wave structured light reveals highly nontrivial features of transverse spin and orbital angular momentum densities. The normalized total angular momentum density is conserved universally, suggesting a mutual conversion of the intrinsic transverse spin and orbital angular momentum in free space. Despite such mutual conversion at local level, the integral intrinsic orbital angular momentum can be well defined with the topological charge of the vortex per photon. Moreover, the orientation of transverse spin density is governed by the direction of the Poynting momentum density, manifesting a spin-momentum-locking effect in free space.
\end{abstract}

\maketitle

\section{Introduction}

Optical angular momentum manifests itself in two distinct forms~\cite{Fran22}. One is spin angular momentum (SAM), which originates from the intrinsic polarization degrees of freedom~\cite{Poyn09}; the other is orbital angular momentum (OAM), which arises from the spatial phase degrees of freedom~\cite{Alle92}. The former is purely intrinsic, or, more specifically, independent of the coordinate origin. The latter is allowed to make a clear distinction between the intrinsic and extrinsic ingredients~\cite{Onei02}. SAM and OAM can be separately observable by the distinct effects upon test particles~\cite{Garc03}. For the conventional light beams, both SAM and OAM are essentially longitudinal since they are aligned with the direction of propagation. 

Nevertheless, there has been a growing interest in the transverse angular momentum of light~\cite{Blin15}. Unlike the conventional and longitudinal angular momentum of light, the optical transverse angular momentum is perpendicular to the propagation direction of light. The transverse SAM (t-SAM) has been generated by evanescent waves~\cite{Blio12,Kiml12,Blio14}, interference fields~\cite{Beks15,Wang18} and focused beams~\cite{Yang11,Math14,Neug15}. The appearance of t-SAM requires the transverse inhomogeneity of the field intensity~\cite{Blin15} and can be closely linked to the curl of the energy flow density~\cite{Shi21,Shi23}. The t-SAM plays crucial roles in the intrinsic spin-momentum-locking effect~\cite{Aiel15,Sdy21} and the intrinsic quantum spin Hall effect of light~\cite{Blis15}. 

The extrinsic transverse OAM (t-OAM) is the most straightforward and trivial form of transverse angular momentum, which is analogous to the classical angular momentum~\cite{Blin15}. Despite its trivial derivation, the extrinsic t-OAM can make remarkable physical effects including the spin-Hall and orbital-Hall effects of light~\cite{Onod04,Blio061,Blio062,Host08,Blio13,Aiel09,Kong12,Korg14,Ling17,Wang19}. Recently, the intrinsic t-OAM is studied intensively in theory~\cite{Blion12,Blio21,Hanc21,Blio23} and in experiment~\cite{Hanc19,Chon20,Guib21,Fang21} with a new type of localized optical pulses characterized by spatiotemporal vortex structure. However, unlike the typical light beams with angular momentum, such pulses are fundamentally polychromatic as opposed to monochromatic.

Thus the intrinsic t-OAM stands for the last missing piece of transverse angular momentum within monochromatic optical fields. Recently, the intrinsic t-OAM of monochromatic optical fields is reported to be generated through the interference of two parallel and counter-propagating linearly-polarized focused beams, and to be observed directly by using an optomechanical sensor~\cite{Huki23}. 

The motivation for this work is to find the intrinsic t-OAM in the monochromatic optical fields. In this work, we propose that the rotating-wave monochromatic structured light can carry the intrinsic t-OAM. In Sec.~\ref{II}, the rotating wave is introduced concisely, which is a monochromatic wave with wave vortex, propagating along the azimuthal direction and consisting of infinite plane waves. In Sec.~\ref{III}, we construct two modes of the rotating-wave structured light in the vector formulation of electromagnetic field. In Sec.~\ref{IV}, we show that the rotating-wave monochromatic structured light can possess an integral intrinsic t-OAM of well-defined value per photon and exhibits locally unusual t-SAM and t-OAM properties. Finally, the conclusion are presented.

\section{Rotating wave}\label{II}
\begin{figure}[ht]
    \centering
    \includegraphics[width = 0.45\columnwidth]
    {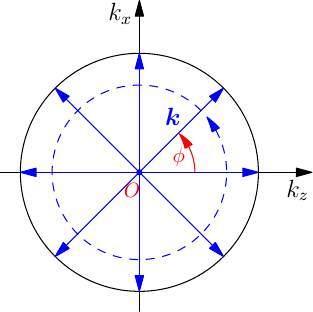} 
    \includegraphics[width = 0.50\columnwidth]
    {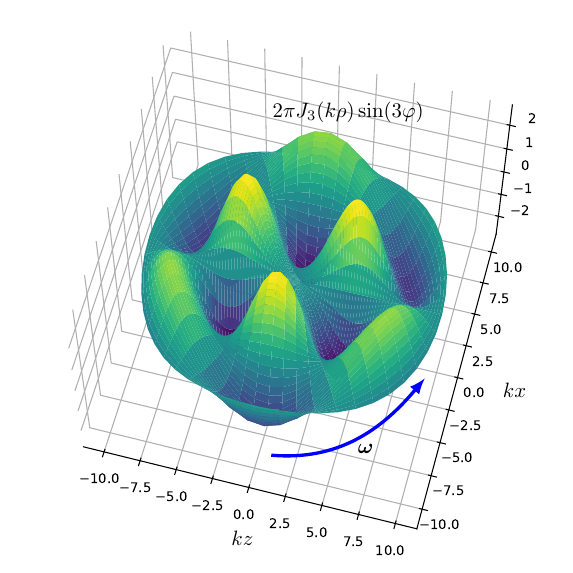} 
    \vskip -0.5 mm    
    \hskip -6.5 mm   (a)  \hskip 35.0 mm (b)
        
    \caption{(a) The wave-vectors of the plane waves point radially outward from the origin and rotates on the $k_y$-axis to form a circle in the plane $(k_z,k_x)$ plane of $\bm{k}$ space. (b) Surface plot of the rotating wave, calculated via the real part of the complex wavefunction in Eq.~\eqref{bes1} with $l=3$. The arrow denotes the wave's direction of propagation.}
    \label{pic1}
\end{figure}
Throughout this paper, let us restrict ourselves to the monochromatic wave field propagating in free space. The typical and simple wave field with OAM is Bessel beam. For simplicity, we can consider specifically a scalar Bessel beam propagating in the $(z,x)$ plane. As illustrated in Fig.~\ref{pic1}(a), such beam can be generated by superposing plane waves with the same frequency $\omega$, wave vectors $\bm k$ pointing radially outward from the origin and forming a circle in the $(k_z,k_x)$ plane, and with the corresponding azimuthal phase difference $l\phi$. Here $l=0,\pm 1,\pm 2,...$ is the topological charge of the vortex and $\phi$ is the azimuthal angle in $\bm k$ space. Thus, the resulting scalar wave function can be written as the integral over $\phi$ from $0$ to $2\pi$:
\begin{align}
\psi(\bm{r},t) &= \int_{0}^{2\pi} e^{ik\cos\phi z + ik\sin\phi x +il\phi-i\omega t} {\rm d}\phi\nonumber\\&=2\pi{i}^lJ_l(k\rho)e^{il\varphi-i\omega t}.\label{bes1}
\end{align}
Here $J_l(k\rho)$ denotes the $l$th-order Bessel function of the first kind, $k=\omega/c$ is the wave number and $c$ is the phase velocity of wave, and $(\rho,\varphi)$ indicate the polar coordinates in the $(z,x)$ plane associated with $z=\rho\cos\varphi$ and $x=\rho\sin\varphi$.

Note that Eq.~\eqref{bes1} stands for a pure rotating wave~\cite{Cepe92}. The term $e^{il\varphi-i\omega t}$ bears resemblance to the typical term $e^{ikx-i\omega t}$ in a traveling wave, enabling the wave to preserve its waveform while propagating along the azimuthal direction in the $(z,x)$ plane. As an illustration, Fig.~\ref{pic1}(b) shows the rotating wave for the mode $l=3$. Unlike traveling waves, rotating waves are confined to specific modes due to their circular propagation. For mechanical waves, such as surface waves in water~\cite{Cepe92} and acoustic waves~\cite{Cepk92,Sant09,Hong15}, rotating waves offer a clear understanding of angular momentum, a concept commonly encountered in quantum waves or electromagnetic waves~\cite{Vela93,Chav96}. More interestingly, rotating waves can provide us with new insights into the intrinsic t-SAM and t-OAM of light, as we will discuss in the following sections.

\section{Rotating-wave structured light}\label{III}

\begin{figure}[ht]
    \centering
    \includegraphics[width = 0.45 \columnwidth]
    {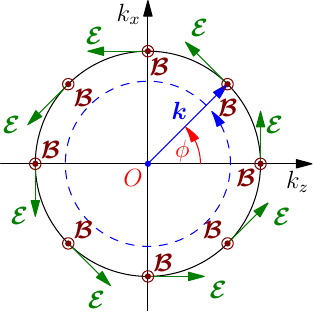}   \hfill\hfill
    \includegraphics[width = 0.45 \columnwidth]
    {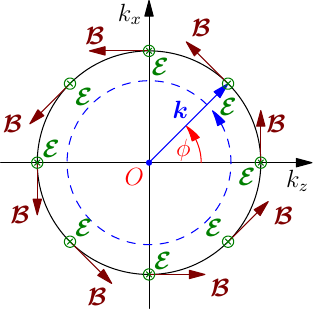} 
    \vskip -0.50 mm    
    \hskip -2.50 mm   (a)  \hskip 42.5 mm (b)
        
    \caption{Schematic diagrams of the two basic modes of polarization. (a) $\bm{\mathcal{B}}$ is polarized in the positive $y$-direction and then $\bm{\mathcal{E}}$ is located in the $(z,x)$ plane. (b) $\bm{\mathcal{E}}$ points in the negative $y$ direction and then $\bm{\mathcal{B}}$ lies in the $(z,x)$ plane. Dots and crosses are used to represent a field coming out of or going into the plane of the paper, respectively.}
    \label{pic2}
\end{figure}

We now consider the rotating-wave structured light. Light is a form of electromagnetic waves which manifest as vector fields, so there is a need for constructing vector beams of rotating wave. For simplicity and more concentration on the intrinsic t-OAM, we only consider the rotating-wave beam consisting entirely of linearly polarized plane waves. Due to the transversality of electromagnetic waves, the electric and magnetic fields $\bm{\mathcal{E}}$, $\bm{\mathcal{B}}$ and the wave vector $\bm k$ of each plane wave in the beam spectrum constitute a right-handed orthogonal triad. As a consequence, two basic modes of polarization should be considered in the problem. One mode is that $\bm{\mathcal{B}}$ is set along the positive $y$-direction and so $\bm{\mathcal{E}}$ is put in the $(z,x)$ plane; see Fig.~\ref{pic2}(a). Conversely, the other one is that  $\bm{\mathcal{E}}$ is fixed along the negative $y$-direction and so $\bm{\mathcal{B}}$ is placed in the $(z,x)$ plane; see Fig.~\ref{pic2}(b).

With the help of Maxwell's equations, the resulting electric and magnetic fields of the first mode can be given by 
\begin{equation}
\bm{B}=\psi(\bm{r},t)\bm{e}_{y},~~~\bm{E}=\frac{i}{k}\na\times\bm{B}.\label{beb}
\end{equation}
Here $\bm{e}_{y}$ is the unit vector in the $y$-direction. For the convenience of future use and with a slight computation cost, the explicit components of the resulting electric field can be obtained
\begin{align}
E_x&=-i^{l+1}\pi\big[J_{l-1}e^{i(l-1)\varphi}-J_{l+1}e^{i(l+1)\varphi}\big]e^{-i\omega t},\label{ex}\\
E_z&=-i^{l}\pi\big[J_{l-1}e^{i(l-1)\varphi}+J_{l+1}e^{i(l+1)\varphi}\big]e^{-i\omega t}.\label{ez}
\end{align}
For notational simplicity, from now on the argument $k\rho$ will be leave out from the Bessel functions. 

In the similar way, the second mode can be represented by
\begin{equation}
\bm{E}=-\psi(\bm{r},t)\bm{e}_{y},~~~\bm{B}=-\frac{i}{k}\na\times\bm{E}.\label{mod2}
\end{equation}
Clearly, the second mode is able to be fulfilled properly from the first one by the exchange $\bm{E}\to \bm{B}$ and $\bm{B}\to -\bm{E}$. In fact, such exchange is a typical transformation of electromagnetic duality in the free space~\cite{Calk65,Jackson}. As a direct result of the transformation of electromagnetic duality, the case of the second mode can be derived from that of the first one. For this reason, we will concentrate on the first mode in the following section.

\section{Spin and orbital angular momenta}\label{IV}

In this core section, we turn now to analyze the t-SAM and t-OAM of the rotating-wave structured light. We start with the time-averaged densities of energy $W$, SAM $\bm S$ and OAM $\bm L$ in a monochromatic optical field with separate electric- and magnetic-field contributions
\begin{eqnarray}
     W&=&W^{e}+W^{m}=\frac{g\omega}{2}\big(|\bm{E}|^2+|\bm{B}|^2\big),\label{web}\\
\bm{S}&=&{\bm S}^{e}+{\bm S}^{m}=\frac{g}{2}\Im\big(\bm{E}^{*}\times\bm{E}+\bm{B}^{*}\times\bm{B}\big),\label{seb}\\
\bm{L}&=&{\bm L}^{e}+{\bm L}^{m}=\frac{g}{2}\Re\big(E^{*}_{j}\hat{\bm{\mathcal L}} E_{j}+B^{*}_{j}\hat{\bm{\mathcal L}} B_{j}\big).\label{leb}
\end{eqnarray}
Here $g=(8\pi\omega)^{-1}$ in Gaussian units, the OAM operator $\hat{\bm{\mathcal L}}=\bm{r}\times \frac{1}{i}\na$ and Einstein summation convention is used for the three-component vector.

Substituting Eqs.~\eqref{beb}-\eqref{ez} into Eqs.~\eqref{web}-\eqref{leb}, we can calculate the time-averaged densities of energy, SAM and OAM. The transverse angular momentum is the focus of attention in this study. Consequently, only the $y$-components of the SAM and OAM densities should be taken into consideration for the light propagating in the $(z,x)$ plane. In fact, the SAM density is totally transverse because it is non-zero solely along the $y$-direction. Omitting the common factor $g\pi^2$ and writing in explicit contributions from electric and magnetic fields, we obtain finally
\begin{align}
W^{e}&=\omega(J_{l-1}^2+J_{l+1}^2),~\qquad\qquad~W^{m}=2\omega J_{l}^2,\label{w-e-b}\\
S^{e}_y&=J_{l-1}^2-J_{l+1}^2,~ ~\qquad~~~~\qquad~S^{m}_{y}=0,\label{s-e-b}\\
L^{e}_y&=(l-1)J_{l-1}^2+(l+1)J_{l+1}^2,~~L^{m}_{y}=2lJ_{l}^{2}.\label{l-e-b}
\end{align}

\begin{figure}[ht]
    \centering
    \includegraphics[width = 1\columnwidth]
    {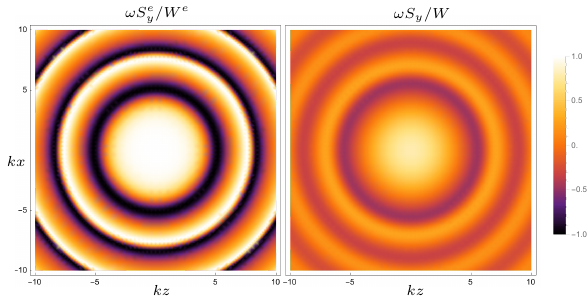}  
     \vskip -0.5 mm 
    \hskip -2.0 mm   (a)  \hskip 32.20 mm (b)
     \vskip 0.50 mm  
    \includegraphics[width = 1\columnwidth]
    {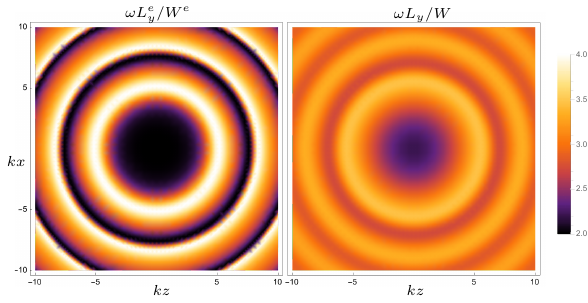}
      \vskip -0.5 mm 
    \hskip -2.0 mm   (c)  \hskip 32.20 mm (d)        
    \caption{Spatial distributions of the normalized t-SAM and t-OAM. (a)~$\omega S_{y}^{e}/W^{e}$, calculated via Eqs.~\eqref{w-e-b} and \eqref{s-e-b}. (b)~$\omega S_{y}/W$, Eqs.~\eqref{Web} and \eqref{Seb}. (c)~$\omega L_{y}^{e}/W^{e}$, Eqs.~\eqref{w-e-b} and \eqref{l-e-b}. (d)~$\omega L_{y}/W$, Eqs.~\eqref{Web} and \eqref{Leb}. The parameter is $l=3$.}
    \label{Tsl} 
\end{figure}

From Eq.~\eqref{s-e-b}, we see that the t-SAM density is completely provided by the roatation of the electric field. The normalized electric SAM density $\omega S_y/W^{e}$ can change sign from $-$ to $+$ and its minimum and maximum values are $-1$ and $1$, respectively; see Fig.~\ref{Tsl}(a). The rotation direction of the electric field then converts from the negative to the positive $\varphi$-direction relative to the $y$-axis. Namely, the electric field can shift from the left-handed to the right-handed circular polarization, and vice versa. For the magnetic sector, we can identify the relation $L_{y}^{m}=lW^{m}/\omega$, and so we can confirm that the t-OAM coincides well with $l\hbar$ per photon as the quantization of photon's energy $\hbar \omega$. However, for the t-OAM density of the electric contribution, there does not exist a similarly direct correlation between $L^{e}_y$ and $W^{e}$. Interestingly, from Eqs.~\eqref{w-e-b}-\eqref{l-e-b} we can find the total transverse angular momentum density of the electric sector satisfies 
\begin{equation}
\mathcal{J}^{e}_{y}=L^{e}_y+S^{e}_{y}=\frac{l}{\omega}W^{e}.\label{Jew}
\end{equation}
Again, for the electric sector the total transverse angular momentum still corresponds to $l\hbar$ per photon. By Eq.~\eqref{Jew}, the normalized electric t-OAM density is then given by $\omega L_y/W^{e}=l-\omega S_y/W^{e}$ and so it can be regarded as varying with $\omega S_y/W^{e}$ if $l$ is given, and vice versa. It achieves its minimum and maximum values of $l-1$ and $l+1$; see Fig.~\ref{Tsl}(c). 

In addition, the densities of energy, t-SAM and t-OAM display strong electric-magnetic asymmetry~\cite{Blin15}, especially the t-SAM density. Not coincidentally, the electric-magnetic asymmetry can be found in the second mode~\eqref{mod2} as well. However, in such case the electric field will make no contribution to the t-SAM density and so only the magnetic field will generate the t-SAM density. In particular, if $l=0$, the optical structure is so trivial that the vortex does not occur and both the densities of SAM and OAM vanish. 

Evidently, the densities of energy, t-SAM and t-OAM contributed by the electric and magnetic fields together can be written as
\begin{align}
W&=\omega(J_{l-1}^2+2J_{l}^2+J_{l+1}^2),\label{Web}\\
S_y&=J_{l-1}^2-J_{l+1}^2,\label{Seb}\\
L_y&=(l-1)J_{l-1}^2+2lJ_{l}^{2}+(l+1)J_{l+1}^2.\label{Leb}
\end{align}
The normalized t-SAM and t-OAM density are depicted in Fig.~\ref{Tsl}(b) and (d). From Eqs.~\eqref{Web}-\eqref{Leb}, we obtain easily the total transverse angular momentum density
\begin{equation}
\mathcal{J}_{y}=L_y+S_{y}=\frac{l}{\omega}W.\label{Jebw}
\end{equation}
Once again, the total transverse angular momentum corresponds to $l\hbar$ per photon. It is natural to interpret Eqs.~\eqref{Jew} and ~\eqref{Jebw} as representing conservation of the transverse angular momentum. Such conservation manifests a new type of mutual conversion of the t-SAM and t-OAM without light-matter interaction~\cite{Sdy21}. Integrating Eq.~\eqref{Jebw} over the $(z,x)$ plane, we obtain the normalized integral values of total transverse angular momentum
\begin{equation}
\frac{\langle\mathcal{J}_{y}\rangle}{\langle W\rangle}=\frac{l}{\omega}.\label{IJ}
\end{equation}
Here $\langle\cdots\rangle$ means the integration over the $(z,x)$ plane, $\langle\cdots\rangle=\int^{\infty}_{-\infty} \int^{\infty}_{-\infty}\cdots{\rm d}z{\rm d}x=\int^{\infty}_{0}\int^{2\pi}_{0} \cdots\rho{\rm d}\rho{\rm d}\varphi$. As seen from Eq.~\eqref{IJ}, the integral total transverse angular momentum can be defined well with $l\hbar$ per photon. More specifically, such integral transverse angular momentum merely stands for the intrinsic t-OAM, which will be elucidated in the remainder of this section.

\begin{figure}[ht]
    \centering
    \includegraphics[width = 0.98 \columnwidth]
    {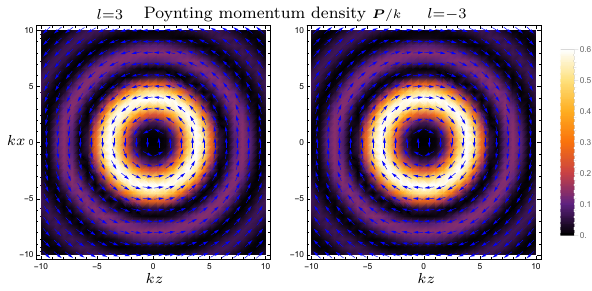}  
    \vskip -0.5 mm 
    \hskip -1.50 mm   (a)  \hskip 33 mm (b)
     \vskip 2.5 mm  
    \includegraphics[width = 0.9\columnwidth]
    {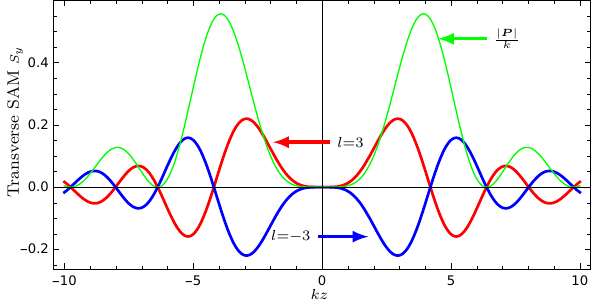}
    \hfill{}
    \vskip -0.5 mm 
    \hskip -2.5 mm  (c)         
    \caption{Spin-momentum locking for the rotating-wave structured light. (a)-(b) The spatial distributions of the Poynting momentum density with $l=\pm 3$, respectively, as given by Eq~\eqref{Poy1}. Arrows are used to indicate the direction of the Poynting momentum density. (c) The spatial distributions of the t-SAM $S_{y}$ along the $z$-axis with $l=\pm 3$, respectively, Eq~\eqref{Seb}.}
    \label{Tsp}
\end{figure}

It is worth further looking into the t-SAM. Following Refs.~\cite{Shi21} and~\cite{Shi23}, the connection between the t-SAM density and the Poynting momentum density for an arbitrary surface evanescent wave can be established as follows:
\begin{equation}
\bm{S}=\frac{1}{2k^2}\na\times\bm{P},\label{sp}
\end{equation}
where the Poynting momentum density takes the form
\begin{equation}
\bm{P}=gk\Re\big({\bm E}^{*}\times {\bm B}\big).\label{Poy}
\end{equation}
Plugging Eqs.~\eqref{beb}-\eqref{ez} into Eq.~\eqref{Poy}, the Poynting momentum density is given by
\begin{equation}
 \bm{P}=\frac{4l}{\rho}J_{l}^2\cos\varphi\bm{e}_x-\frac{4l}{\rho}J_{l}^2\sin\varphi\bm{e}_z=\frac{4l}{\rho}J_{l}^2\bm{e}_{\varphi}.\label{Poy1}
\end{equation} 
Here the common factor $g\pi^2$ has been omitted. From Eq.~\eqref{Poy1}, we find that the Poynting momentum density is directed along purely the azimuthal direction in the $(z,x)$ plane, as illustrated in Fig.~\ref{Tsp}(a)-(b). Therefore, the $y$-components of SAM and OAM are indeed transverse because of their orientations perpendicular to the Poynting momentum density as well as the propagation direction. 

With elementary algebra, we can verify that the connection~\eqref{sp} still holds for the t-SAM density~\eqref{Seb} and the Poynting momentum~\eqref{Poy} despite non-evanescent wave. Such connection offers a guideline for understanding the t-SAM from the perspective of the transverse gradient of Poynting momentum. The t-SAM here is the interference result of plane waves with different wave-vector directions in the same plane, even taking place in the very simple interference systems of two~\cite{Beks15} or three~\cite{Wang18} plane waves. In our instance here, the t-SAM density reversals the local orientations universally once the Poynting momentum changes from the positive to the negative $\varphi$-direction, as illustrated in Fig.~\ref{Tsp}(c). Thus, the direction of the t-SAM density is determined by the direction of Poynting momentum density, which can be derived clearly from the connection~\eqref{sp}. From the connection between the t-SAM density and the Poynting momentum, we show that the rotating-wave structured light reveals an intrinsic spin-momentum-locking effect~\cite{Aiel15,Sdy21} without light-matter interaction.

Let us coming back to the integral transverse angular momentum. In general, the t-SAM of helicity independence exists locally, and yet its integral value vanishes~\cite{Blin15}. The integral t-SAM in Eq.~\eqref{Seb} can be verified to go to exactly zero:
\begin{align}
\langle S_{y}\rangle &=\int^{2\pi}_{0}{\rm d}\varphi\int^{\infty}_{0}(J_{l-1}^2-J_{l+1}^2)\rho{\rm d}\rho\nn\\
                     &=\frac{8\pi l}{k^2}\int^{\infty}_{0}J_{l}(k\rho)\frac{{\rm d}J_{l}(k\rho)}{{\rm d}(k\rho)}{\rm d}(k\rho)=0.
\end{align}
Then the integral total transverse angular momentum can be written as 
\begin{equation}
\langle\mathcal{J}_{y}\rangle=\langle L_{y}\rangle+\langle S_{y}\rangle=\langle L_{y}\rangle=\frac{l}{\omega}\langle W\rangle. \label{lw}
\end{equation}
Hence the integral transverse angular momentum belongs entirely to the t-OAM. 

An alternate approach to derive the above result is giving by performing directly the integrations of energy $\langle W^{e}\rangle $ and $\langle W^{m}\rangle $ and the t-OAM $\langle L^{e}_{y}\rangle $ and $\langle L^{m}_{y}\rangle $. Using the closure equation of Bessel Functions of the first kind~\cite{Arfken}, we have
\begin{equation}
\langle W^{e}\rangle=\langle W^{m}\rangle,~~\langle L^{e}_{y}\rangle=\langle L^{m}_{y}\rangle.
\end{equation}
Immediately, we obtain~\cite{note}
\begin{equation}
\frac{\langle L_{y}\rangle}{\langle W\rangle}=\frac{\langle L^{e}_{y}\rangle}{\langle W^{e}\rangle}=\frac{\langle L^{m}_{y}\rangle}{\langle W^{m}\rangle}=\frac{l}{\omega}.
\end{equation}
Thus, the relation~\eqref{lw} between the integral t-OAM and energy is recovered, and the integral t-SAM vanishes.

We are now in the position to interpret why the integral t-OAM is intrinsic. For a translation of the coordinate origin $\bm{r}\to \bm{r}+\bm{r}_{0}$, the integral t-OAM is changed by $\langle L_{y}\rangle \to \langle L_{y}\rangle +\langle L_{y}^{\rm ext}\rangle=\langle L_{y}\rangle +\big(\bm{r}_{0}\times \langle \bm{P} \rangle\big)_y$. From Eq.~\eqref{Poy1} we find $\langle \bm{P} \rangle=0$ and so the extrinsic t-OAM $\langle L_{y}^{\rm ext}\rangle$ is equal to zero. The integral t-OAM is unchanged under the translation of the coordinate origin and therefore it is intrinsic. 

\section{Discussion and Summary}
In summary, we have shown that the rotating-wave structured light carries purely integral intrinsic t-OAM within monochromatic optical fields. Furthermore, the rotating-wave structured light with wave vortex, consisting of linearly-polarized plane waves, exhibits locally unique t-SAM and t-OAM features. The sum of the locally-normalized t-SAM and t-OAM is conserved universally, which manifests a strong coupling of t-SAM and t-OAM. The t-SAM and t-OAM variations are heavily dependent on each other. Such strong coupling uncovers a new type of mutual conversion of the intrinsic t-SAM and t-OAM without light-matter interaction. In spite of such locally mutual conversion, the integral t-SAM and t-OAM are rather robust, and the integral intrinsic transverse angular momentum belong purely to the intrinsic t-OAM. 

Impressively, the t-SAM density of the rotating-wave structured light also can be directly linked to the curl of the Poynting momentum density even though it is not associated with evanescent waves. As a consequence, the orientation of the local t-SAM is made directly subject to the local direction of the Poynting momentum density, which can be recognized as an intrinsic spin-momentum-locking effect of the rotating-wave structured light in free space.

We look forward to the future experimental verification of such nontrivial properties of the spin and orbital angular momentum in the rotating-wave structured light. The rotating-wave structured light might offer high potential for applications in the optical manipulation of matter.

\section*{Acknowledgments}
The author is grateful to Z. Xiong and Y. Liu for fruitful discussions. This work was funded by the Scientific Research Project of Hubei Polytechnic University (No. 20xjz02R).


\begin{thebibliography}{99}

\bibitem{Fran22} S. Franke-Arnold, Nat. Rev. Phys. {\bf 4}, 361 (2022).
\bibitem{Poyn09} J.H. Poynting, Proc. R. Soc. Lond. A {\bf 82}, 560 (1909).
\bibitem{Alle92} L. Allen, M.W. Beijersbergen, R.J.C. Spreeuw, and J.P. Woerdman, Phys. Rev. A {\bf 45}, 8185 (1992).
\bibitem{Onei02} A.T. O'Neil, I. MacVicar, L. Allen, and M.J. Padgett, Phys. Rev. Lett. {\bf 88}, 053601 (2002).
\bibitem{Garc03}  V. Garc\'es-Ch\'avez, D. McGloin, M.J. Padgett, W. Dultz, H. Schmitzer, and K. Dholakia, Phys. Rev. Lett. {\bf 91}, 093602 (2003).

\bibitem{Blin15} K.Y. Bliokh and F. Nori, Phys. Rep. {\bf 592}, 1 (2015).
\bibitem{Blio12} K.Y. Bliokh and F. Nori, Phys. Rev. A {\bf 85}, 061801 (2012).  
\bibitem{Kiml12} K.-Y. Kim, I.-M. Lee, J. Kim, J. Jung, and B. Lee, Phys. Rev. A {\bf 86}, 063805 (2012).
\bibitem{Blio14} K.Y. Bliokh, A.Y. Bekshaev, and F. Nori, Nat. Commun. {\bf 5}, 3300 (2014).

\bibitem{Beks15} A.Y. Bekshaev, K.Y. Bliokh, and F. Nori, Phys. Rev. X {\bf 5}, 011039 (2015).
\bibitem{Wang18} Z.-L. Wang, D.-D. Lian, and X.-S. Chen, Opt. Express {\bf 26}, 33712 (2018).
\bibitem{Yang11} N. Yang and A.E. Cohen, J. Phys. Chem. B {\bf 115}, 5304 (2011).
\bibitem{Math14} R. Mathevet and G.L.J.A. Rikken, Opt. Mater. Express {\bf 4}, 2574 (2014).
\bibitem{Neug15} M. Neugebauer, T. Bauer, A. Aiello, and P. Banzer, Phys. Rev. Lett. {\bf 114}, 063901 (2015).
 
\bibitem{Shi21} P. Shi, L. Du, C. Li, A.V. Zayats, and X. Yuan, Proc. Natl. Acad. Sci. U.S.A. {\bf 118}, e2018816118 (2021).
\bibitem{Shi23} P. Shi, L. Du, A. Yang, X. Yin, X. Lei, and X. Yuan, Commun. Phys. {\bf 6}, 283 (2023).

\bibitem{Aiel15} A. Aiello, P. Banzer, M. Neugebauer, and G. Leuchs, Nat. Photonics {\bf 9}, 789 (2015).
\bibitem{Sdy21} P. Shi, L. Du, and X. Yuan,  Nanophotonics {\bf 10}, 3927 (2021).
\bibitem{Blis15} K.Y. Bliokh, D. Smirnova, and F. Nori, Science {\bf 348}, 1448 (2015).

\bibitem{Onod04} M. Onoda, S. Murakami, and N. Nagaosa, Phys. Rev. Lett. {\bf 93}, 083901 (2004).
\bibitem{Blio061} K.Y. Bliokh and Y.P. Bliokh, Phys. Rev. Lett. {\bf 96}, 073903 (2006).
\bibitem{Blio062} K.Y. Bliokh, Phys. Rev. Lett. {\bf 97}, 043901 (2006).
\bibitem{Host08} O. Hosten and P. Kwiat, Science {\bf 319}, 787 (2008).
\bibitem{Blio13} K.Y. Bliokh and A. Aiello, J. Opt. {\bf 15}, 014001 (2013)
\bibitem{Aiel09} A. Aiello, N. Lindlein, C. Marquardt, and G. Leuchs, Phys. Rev. Lett. {\bf 103}, 100401 (2009).
\bibitem{Kong12} L.-J. Kong, S.-X. Qian, Z.-C. Ren, X.-L. Wang, and H.-T. Wang, Phys. Rev. A {\bf 85}, 035804 (2012).
\bibitem{Korg14}J. Korger, A. Aiello, V. Chille, P. Banzer, C. Wittmann, N. Lindlein, C. Marquardt, and G. Leuchs, Phys. Rev. Lett. {\bf 112}, 113902 (2014).
\bibitem{Ling17} X. Ling, X. Zhou, K. Huang, Y. Liu, C.-W. Qiu, H. Luo, and S. Wen, Rep. Prog. Phys. {\bf 80}, 066401 (2017).
\bibitem{Wang19} Z.-L. Wang and X.-S. Chen, Phys. Rev. A {\bf 99}, 063832 (2019).

\bibitem{Blion12} K.Y. Bliokh and F. Nori, Phys. Rev. A {\bf 86}, 033824 (2012).
\bibitem{Blio21} K.Y. Bliokh, Phys. Rev. Lett. {\bf 126}, 243601 (2021).
\bibitem{Hanc21} S.W. Hancock, S. Zahedpour, and H.M. Milchberg, Phys. Rev. Lett. {\bf 127}, 193901 (2021).
\bibitem{Blio23} K.Y. Bliokh, Phys. Rev. A {\bf 107}, L031501 (2023).


\bibitem{Hanc19} S.W. Hancock, S. Zahedpour, A. Goffin, and H.M. Milchberg, Optica {\bf 6}, 1547 (2019).
\bibitem{Chon20} A. Chong, C. Wan, J. Chen, and Q. Zhan, Nat. Photonics {\bf 14}, 350 (2020).
\bibitem{Guib21} G. Gui, N.J. Brooks, H.C. Kapteyn, M.M. Murnane, and C.T. Liao, Nat. Photonics {\bf 15}, 608 (2021).
\bibitem{Fang21} Y. Fang, S. Lu, and Y. Liu, Phys. Rev. Lett. {\bf 127}, 273901 (2021).

\bibitem{Huki23} Y. Hu, J.J. Kingsley-Smith, M. Nikkhou, J.A. Sabin, F.J. Rodr\'iguez-Fortu\~no, X. Xu, and J. Millen, Nat. Commun. {\bf 14}, 2638 (2023).

\bibitem{Cepe92} P.H. Ceperley, Rotating Waves, Am. J. Phys. {\bf 60}, 938 (1992).

\bibitem{Cepk92} P.H. Ceperley and A. Koren, J. Acoust. Soc. Am. {\bf 91}, 2331 (1992).
\bibitem{Sant09} A.O. Santill\'an and K. Volke-Sep\'ulveda,  Am. J. Phys. {\bf 77}, 209 (2009).
\bibitem{Hong15} Z. Hong, J. Zhang, and B.W. Drinkwater, Phys. Rev. Lett. {\bf 114}, 214301 (2015).

\bibitem{Vela93} J.E. Velazco and P.H. Ceperley, IEEE Trans. Microwave Theory Techn. {\bf 41}, 330 (1993).
\bibitem{Chav96} S. Ch\'avez-Cerda, G.S. McDonald, and G.H.C. New, Opt. Commun. {\bf 123}, 225 (1996).

\bibitem{Calk65} M.G. Calkin, Am. J. Phys. {\bf 33}, 958 (1965).
\bibitem{Jackson} J.D. Jackson, {\it Classical electrodynamics}, 3rd ed. (Wiley, Hoboken, NJ, 1999). 

\bibitem{Arfken} G.B. Arfken and H.J. Weber, {\it Mathematical methods for physicists}, 6th ed. (Elsevier Academic Press, New York, 2005).
\bibitem{note} Despite $\langle W^{e}\rangle =\langle W^{m}\rangle$ and $\langle L^{e}_{y}\rangle=\langle L^{m}_{y}\rangle$ are infinite, they indeed can be rigorously defined by the Dirac $\delta$-function, and the normalized value of integral t-OAM $\langle L_{y}\rangle/\langle W\rangle$ is exactly $l/\omega$. 


\end{thebibliography}
\end{document}